\documentclass[review]{elsarticle}

\graphicspath{{./figures/}}  

\usepackage{graphicx}
\usepackage{hyperref}
\usepackage{float}
\usepackage{amsmath}
\usepackage{amsfonts}
\usepackage{algorithm}
\usepackage{rotating}
\usepackage{booktabs} 
\usepackage{cleveref}
\usepackage{algpseudocode}
\usepackage{verbatim} 
\usepackage{apalike}
\usepackage{booktabs}
\usepackage{tabularx} 
\usepackage[left=2.5cm,right=2.5cm]{geometry}
\usepackage{tikz}
\usetikzlibrary{arrows.meta, positioning, decorations.pathmorphing, calc}
\restylefloat{table}

\journal{Expert Systems with Applications}

\biboptions{authoryear}

\begin{document}
\begin{frontmatter}


\begin{titlepage}
\begin{center}
\vspace*{1cm}

\textbf{ \large The Enhanced Physics-Informed Kolmogorov–Arnold Networks: Applications of Newton's Laws in Financial Deep Reinforcement Learning Algorithms }

\vspace{1.5cm}

Trang Thoi$^{a,*}$ (tdktrang@vt.edu), Hung Tran$^{b,*}$ (hungqtran@vt.edu), Tram Thoi$^c$ (tramthoi.81@gmail.com),  Huaiyang Zhong$^{d}$ (hzhong@vt.edu)\\

\hspace{10pt}

\begin{flushleft}
\small  
$^a$ Virginia Polytechnic Institute and State University, Blacksburg, USA \\
$^b$ Virginia Polytechnic Institute and State University, Blacksburg, USA \\
$^c$ Ho Chi Minh University of Banking, Ho Chi Minh City, Vietnam\\
$^d$ Virginia Polytechnic Institute and State University, Blacksburg, USA 

\vspace{1cm}
\textbf{Corresponding author at: Virginia Polytechnic Institute and State University, Blacksburg, USA } \\

Trang Thoi \\
Virginia Polytechnic Institute and State University, Blacksburg, USA \\
Email: tdktrang@vt.edu

Hung Tran \\
Virginia Polytechnic Institute and State University, Blacksburg, USA \\
Email: hungqtran@vt.edu

\end{flushleft}        
\end{center}
\end{titlepage}

\title{The Enhanced Physics-Informed Kolmogorov–Arnold Networks: Applications of Newton's Laws in Financial Deep Reinforcement Learning Algorithms}

\author[label1]{Trang Thoi\corref{cor1}}
\ead{tdktrang@vt.edu}

\author[label2]{Hung Tran\corref{cor1}}
\ead{hungqtran@vt.edu}

\author[label3]{Tram Thoi}
\ead{tramthoi.81@gmail.com}

\author[label4]{Huaiyang Zhong}
\ead{hzhong@vt.edu}

\cortext[cor1]{Corresponding author.}
\address[label1]{Virginia Polytechnic Institute and State University, Blacksburg, USA}
\address[label2]{Virginia Polytechnic Institute and State University, Blacksburg, USA}
\address[label3]{Ho Chi Minh University of Banking, Ho Chi Minh City, Vietnam}
\address[label4]{Virginia Polytechnic Institute and State University, Blacksburg, USA}

\begin{abstract}

Under the field of quantitative finance, Deep Reinforcement Learning (DRL) is commonly used either to generate discrete trade signals or to determine continuous portfolio allocations. In this work, we propose a novel reinforcement learning framework for portfolio optimization that incorporates Physics-Informed Kolmogorov-Arnold Networks (PIKANs) into several DRL algorithms. The approach replaces conventional multilayer perceptrons with Kolmogorov-Arnold Networks (KANs) in both actor and critic components—utilizing learnable B-spline univariate functions to achieve parameter-efficient and more interpretable function approximation. During actor updates, we introduce a physics-informed regularization loss that promotes second-order temporal consistency between observed return dynamics and the action-induced portfolio adjustments. The proposed framework is evaluated across three equity markets—the United States, Vietnam, and China, covering both emerging and developed economies. Across all three markets, PIKAN-based agents consistently deliver higher cumulative and annualized returns, superior Sharpe and Calmar ratios, and more favorable drawdown characteristics compared to both standard DRL baselines and classical online portfolio-selection methods. This yields more stable training, higher Sharpe ratios, and superior performance compared to traditional DRL counterparts. The approach is particularly valuable in highly dynamic and noisy financial markets, where conventional DRL often suffers from instability and poor generalization.


\end{abstract}


\end{frontmatter}

\section{Introduction}
\label{introduction}

In the last several years, online learning algorithms have gained significant momentum from the success of deep learning methodologies and the rise of deep reinforcement learning (DRL). DRL merges deep neural networks with reinforcement learning (RL), enabling agents to solve complex sequential decision-making problems through interaction with dynamic environments. This capability makes DRL particularly impactful in areas such as algorithmic trading and portfolio management.In this study, we propose innovative approaches to incorporating a physics-inspired regularization framework motivated by second-order temporal dynamics into traditional RL neural networks. Drawing inspiration from Physics-Informed Neural Networks (PINNs), which integrate known dynamics directly into the loss function to enhance physical plausibility, data efficiency, and generalization \citep{cuomo_scientific_2022, karniadakis_physics-informed_2021, banerjee_survey_2025}, we adapt this concept to financial time series. Notably, hybrid approaches combining PINNs with Kolmogorov-Arnold Networks (KANs) have demonstrated remarkable accuracy and generalizability by flexibly embedding domain knowledge into deep networks \citep{zhang_physics-informed_2025}. These advances highlight the potential of physics-informed methods to bridge theoretical priors and data-driven learning in complex, noisy domains.

By extension, our work bridges several gaps in the literature on financial reinforcement learning, where most models rely mainly on data-driven signals and ignore the structural constraints that can regularize learning. To adapt to the complex physics loss calculation, we add a physics-informed loss that compares observed market velocities and accelerations with those predicted by a force-based formulation of Newton's Second Law, and directly integrate Newtonian motion principles into the agent's update step. The agent uses adaptive weighting to preserve stability while optimizing this physics loss and the standard RL objective during training. In the project, we also incorporate the Kolmogorov-Arnold Network (KAN) architecture to support the physical loss mechanism, allowing for a flexible representation of nonlinear financial dynamics through spline-based functional mappings. Our improved A2C, DDPG, PPO, and TD3 agents learn decision policies that represent market momentum and dynamic structure by incorporating physical consistency into the learning loop. This results in more stable learning and higher Sharpe Ratios than their traditional DRL counterparts. This direction is important because financial markets are extremely dynamic and noisy, and conventional DRL models often struggle with instability and poor generalization. By integrating Newtonian principles into both the network architecture and training objective, our method aims to generate more resilient, comprehensible, and consistently performing portfolio strategies in the face of shifting market conditions.

We clarify that the Newtonian formulation employed here is not proposed as a literal physical model for financial markets. It rather serves as a structured inductive bias designed to impose second-order temporal consistency on the learning dynamics. By treating asset returns as analogous to velocity-like quantities and penalizing substantial deviations between empirically observed return changes and the portfolio adjustments implied by the agent's actions, the regularization term effectively mitigates abrupt reallocation attributable to transient noise. While akin to regularization techniques that promote action smoothness or momentum preservation, this approach distinguishes itself by explicitly coupling the agent's action updates to the underlying return dynamics, rather than applying solely exogenous smoothing penalties.

\section{Related Work}
\subsection{Machine Learning in Portfolio Optimization}
Intelligent approaches to portfolio optimization have been increasingly adopted due to their ability to learn from and analyze historical data in volatile market conditions \citep{alpaydin_introduction_2020,gunjan_brief_2023}. A common application is the integration with classical frameworks such as the Mean–Variance (MV) model proposed by \citet{markowitz_portfolio_2008}, which quantifies returns and risks by means and variances, allowing investors to balance expected return against risk; for instance,\citet{chen_meanvariance_2021} combined a hybrid model named IFAXGBoost to predict potential stocks with higher returns, which were then incorporated into an MV framework for portfolio selection. Similarly, \citet{du_meanvariance_2022} demonstrated an attention-based LSTM combined with the MV model enhances spread return prediction and cointegrated stock pair selection, producing more profitable portfolios than conventional ML approaches. \citet{min_robust_2021} addressed the conservatism of worst-case robust portfolio optimization by proposing hybrid robust models under ellipsoidal uncertainty sets. Using LSTM and XGBoost to forecast market movements, they developed the hybrid portfolios based on variance (HRMV), showing that HRMV outperforms the traditional MV. These studies highlight the relationship that enhancing the prediction of future stock market performance also contributes positively to portfolio construction.

\subsection{Deep Reinforcement Learning}
Deep Reinforcement Learning (DRL), a branch of machine learning concerned with sequential decision-making, has recently emerged as a compelling approach for financial trading tasks. Its applications in this domain mainly fall into two categories: generating discrete trading signals or producing continuous portfolio weights. Regarding the former, the Deep Q-Network (DQN) introduced by \citet{mnih_human-level_2015} and Double Deep Q-Network (DDQN) introduced by \citet{hasselt_deep_2016}, as value-based deep reinforcement learning methods, have been widely adopted to determine buy, hold, and sell actions \citet{liu_practical_2018, jeong_improving_2019,le_thi_reinforcement_2020,bajpai_application_2021,shi_stock_2021,zejnullahu_applications_2022} (Liu et al., 2018; Xiong et al., 2018; Chen and Gao, 2019; Jeong and Kim, 2019; Dang, 2020; Bajpai, 2021; Shi et al., 2021; Zejnullahu et al., 2022). Other advanced versions, TDQN \citep{theate_application_2021} was inspired by DQN to optimize trading positions by maximizing the Sharpe ratio, using synthetic trajectories for training and a rigorous evaluation protocol. Multi-DQN \citep{carta_multi-dqn_2021}, an ensemble of Q-learning agents trained multiple times on the same data, learnt to maximize returns without labeled market movements and derives multi-asset trading strategies. However, DQN struggles to handle continuous state and action spaces, which limits its applicability to portfolio optimization problems where actions correspond to portfolio weights.
In contrast with critic-only methods such as DQN, actor-critic DRL approaches can be a good solution for portfolio problems. Building on this view, \citet{rezaei_taxonomy_2025} Rezaei and Nezamabadi-Pour (2025) provided a comprehensive synthesis of DRL applications in portfolio optimization and contend that actor-critic methods (A2C, DDPG, PPO, SAC, TD3) are better suited to continuous portfolio weights than DQN. \citet{liang_adversarial_2018} Liang et al. (2018) benchmarked DDPG, PPO and PG on Chinese equities and introduced adversarial training by injecting noise into prices, which raises average daily return and Sharpe ratio, an evidence that robust training can materially improve DRL portfolio agents. \citet{durall_asset_2022} linked classical allocation to modern DRL, modeling actions as continuous portfolio weights and surveying stability fixes such as TD3 (to curb Q-overestimation) and SAC (entropy-regularized exploration). \citet{zhang_cost-sensitive_2020} moved beyond return-only rewards by making portfolio RL explicitly cost-sensitive, integrating transaction costs into the objective. In compliment, \cite{jiang_deep_2024} designed a DRL-TD3-based risk model for portfolio selection and highlighted the role of risk aversion and trading frictions in contemporary designs. More recent technique papers refine objectives and representations: \citet{sattar_novel_2025} propose an RMS-driven DRL framework to balance reward and risk in stock trading, reporting improved profitability and risk control, while \citet{sang_portfolio_2025} introduce MT-Exp, which pairs multi-task self-supervised representation learning with adaptive exploration to boost sample efficiency and generalization in portfolio management. 

\subsection{Physics-Informed Neural Network Design}

Existing studies remain fragile due to non-stationary and noisy price dynamics, which weaken effective representation learning. These architectures cannot frequently jointly capture temporal dynamics and inter-asset correlations, and thus generalize poorly under concept drift and simulation-to-market gaps, leading to unstable and weak out-of-sample performance. Physics-informed neural networks are a relatively underexplored approach that can address these limitations by improving learning in dynamic financial and complex optimization contexts. The networks represent a paradigm in scientific machine learning that integrates physical laws, typically expressed as partial differential equations (PDEs), into neural network architectures to solve forward and inverse problems with sparse or noisy data. The foundational concept emerged from early works on data-driven solutions for nonlinear PDEs, where neural networks are trained to respect governing equations via a composite loss function combining data fidelity and physics residuals. \citet{raissi_physics-informed_2019} formalized PINNs as a multi-task learning framework, encoding PDEs (Burgers' equation) as soft constraints in the loss, enabling unsupervised training without labeled solutions. This approach contrasts with traditional numerical methods like finite elements, offering mesh-free flexibility and handling high-dimensional problems more efficiently. \citet{noguer_i_alonso_physics-informed_2023} showed that PINNs embed Black-Scholes and Heston dynamics directly into training, producing predictions that are both data-consistent and aligned with established financial rules and stochastic differential equations. Similarly, \citet{zideh_physics-informed_2024} emphasized that physics-informed machine learning enhances the reliability, interpretability, and feasibility of data-driven tools for power grids by embedding physical constraints, thereby overcoming key limitations of purely model-based or data-driven methods. \citet{seo_solving_2024} further showed that physics-informed neural networks can efficiently solve real-world optimization tasks by integrating governing laws, constraints, and goals into the learning process, enabling top-down searches that require less exploration and successfully capturing narrow or unstable solutions beyond the reach of traditional reinforcement learning. Extending this perspective, \citet{mortezanejad_physics-informed_2025} proposed integrating physical equations into neural frameworks for multivariate time series forecasting, demonstrating that embedding physics as soft constraints improves accuracy, robustness, and uncertainty quantification even with sparse or noisy data.

In the recent years, there are several innovative approaches in redesigning traditional neural networks to adapt with the physic-informed structure as well as mathematical uncertainty. Kolmogorov-Arnold Networks (KANs), introduced in 2024, decompose multivariate functions into sums of learnable univariate splines or wavelets, offering parameter efficiency, reduced overfitting, and superior interpretability compared to multilayer perceptrons (MLPs). Physics-Informed Kolmogorov-Arnold Networks (PIKANs) extend PINNs by replacing MLPs with KANs, enhancing expressivity while enforcing physical constraints for PDE solutions. \citet{shuai_physics-informed_2025} apply PIKANs to power system dynamics, achieving lower errors in parameter identification and rotor angle prediction. In reservoir engineering, PIKANs solve Buckley-Leverett equations with high computational efficiency, marking their debut in fluid flow modeling \citep{rao_physics-informed_2025}. Heterogeneous media flow problems benefit from PIKAN's mixed pressure-velocity formulations, demonstrating improved accuracy over traditional PINNs \citep{rao_physics-informed_2025-1}. Efficient implementations like efficient-KAN and WAV-KAN enable dynamical analysis of ODEs/PDEs with unsupervised and supervised methods, achieving up to 99\% benchmark accuracy \citep{patra_physics_2024}. PIKAN variants also address scalability and robustness issues. Separable PIKANs (SPIKANs) apply variable separation to high-dimensional PDEs, reducing training complexity while maintaining or exceeding accuracy \citep{jacob_spikans_2024}. Kolmogorov-Arnold-Informed Neural Networks (KINN) tackle forward and inverse problems across strong, energy, and inverse PDE forms, validating performance in engineering mechanics \citep{wang_kolmogorovarnold-informed_2025}. \citet{faroughi_neural_2025} also introduce a provocative approach, Chebyshev-based PIKANs (cPIKANs), to enhance noise resilience and spectral control, with neural tangent kernel analyses probing convergence against PINNs. Other designs, such as the modified architectures like tanh-cPIKAN, introduce nonlinearities for stability in pharmacology gray-box discovery or Bayesian frameworks in PIKANs handle uncertainty in noisy data scenarios \citep{daryakenari_representation_2025,toscano_pinns_2024}.

\section{Problem Formulation}

Active portfolio management is a sequential decision-making problem that naturally fits into a Reinforcement Learning (RL) formulation \cite{yu_model-based_2019}. 
We model it as a finite-horizon, discrete-time Markov decision process (MDP), represented by a tuple 
$(\mathcal{S}, \mathcal{A}, P, R, \gamma)$, where $\mathcal{S}$ is the state space, $\mathcal{A}$ the action space, 
$P$ the transition probability, $R$ the reward function, and $\gamma \in [0,1)$ the discount factor.

\subsection{State Space}

The state $s_t \in \mathcal{S}$ at time $t$ provides the agent with market information required for portfolio allocation. 
Following recent financial RL research \cite{jang_deep_2023, park_novel_2024, betancourt_deep_2021}, the state contains exclusively historical market features and does not include the agent’s current portfolio weights.

Formally, the state is represented as
\begin{align}
    s_t = X_t,
    \label{eq:statespace}
\end{align}
where $X_t$ is a tensor of size $(w \times m \times n)$, containing a lookback window of $w$ trading days for $m$ assets and $n$ features per asset. 
We set $w = 5$.

The feature set consists of daily OHLCV data together with a comprehensive set of twelve widely used technical indicators capturing price trend, volatility, momentum, trading volume, and market strength. Table~\ref{tab:tech_indicators} lists the indicators used to construct the input tensor $X_t$.

\begin{table}[t]
\caption{Technical indicators used as input-state features.}
\label{tab:tech_indicators}
\centering
\begin{tabular*}{\textwidth}{@{\extracolsep{\fill}} l l l}
\toprule
\textbf{Indicator} & \textbf{Description} & \textbf{Parameters} \\
\midrule
ADX     & Average Directional Index                 & timeperiod = 14 \\
ATR     & Average True Range                        & timeperiod = 14 \\
BBANDS  & Bollinger Bands                           & timeperiod = 20,\; $k = 2$ \\
MACD    & Moving Average Convergence/Divergence     & fast=12,\; slow=26 \\
MACDS   & MACD Signal (EMA)                         & signalperiod = 9 \\
MOM     & Momentum                                  & timeperiod = 10 \\
OBV     & On-Balance Volume                         & N/A \\
RET     & Price Return                              & horizons = 1,\; 5 \\
RSI     & Relative Strength Index                   & timeperiod = 14 \\
RVOL    & Realized Volatility                       & timeperiod = 20 \\
WILLR   & Williams’ \%R                             & timeperiod = 60 \\
\bottomrule
\end{tabular*}
\end{table}

\subsection{Action Space}

The action $a_t \in \mathcal{A}$ represents the portfolio rebalancing decision at time $t$, encoded as a continuous-valued weight vector:
\begin{align}
    \mathbf{w}_t = [w_{1,t}, w_{2,t}, \dots, w_{m,t}]^\top.
    \label{eq:weight_vector}
\end{align}
The agent allocates capital across $m$ risky assets under the constraints:
\begin{align}
\sum_{i=1}^{m} w_{i,t} = 1, 
\quad 
w_{i,t} \ge 0,\;\forall i.
\label{eq:weight_total}
\end{align}
This continuous action setting is compatible with modern policy-gradient and actor–critic algorithms such as PPO and DDPG \cite{jang_deep_2023, sun_graphsage_2024}.

\subsection{Reward Function}

The reward $r_t$ evaluates the agent’s portfolio performance from the previous allocation 
$\mathbf{w}_{t-1}$ during period $t$, including transaction costs. 
Following \cite{yu_model-based_2019} and standard financial modeling practice \cite{qiu2024design}, we use the \emph{logarithmic net return}:
\begin{align}
    r_t 
    = \ln\!\bigl(\mu_t \cdot (\mathbf{w}_{t-1}^\top \mathbf{y}_t)\bigr),
    \label{eq:return_reward}
\end{align}
where $\mathbf{y}_t$ is the price-relative vector and $\mu_t$ is the transaction-cost factor defined in \eqref{eq:net_of_fee_factor}.

The agent learns a policy $\pi_\theta(a_t \mid s_t)$ to maximize the expected discounted cumulative reward:
\begin{align}
    J(\theta) 
    = \mathbb{E}_{\tau \sim \pi_\theta} 
    \left[ 
        \sum_{t=0}^{T} \gamma^t r_t 
    \right].
    \label{eq:agent_objectives}
\end{align}

\section{Trading Procedure}
\label{sec:trading_procedure}

The trading mechanism follows the procedure in \cite{yu_model-based_2019}. 
At the beginning of period $t$, the portfolio has value $P_{t-1}$ and weight vector $\mathbf{w}_{t-1}$. 
Asset prices evolve according to the price-relative vector:
\begin{align}
    \mathbf{y}_t
    =
    \left[
        \frac{v_{1,t}}{v_{1,t-1}},\,
        \frac{v_{2,t}}{v_{2,t-1}},\,
        \dots,\,
        \frac{v_{m,t}}{v_{m,t-1}}
    \right]^{\!\top},
    \label{price_vector}
\end{align}
where $v_{i,t}$ denotes the closing price of asset $i$ at time $t$.

\paragraph{Post-price drift.}
After price movement but before rebalancing, the weights drift according to
\begin{align}
    \bar{\mathbf{w}}_t
    = 
    \frac{\mathbf{y}_t \odot \mathbf{w}_{t-1}}
         {\mathbf{w}_{t-1}^\top \mathbf{y}_t},
    \label{eq:drift_weights}
\end{align}
where $\odot$ denotes element-wise multiplication.

\paragraph{Transaction costs.}
The agent then chooses new target weights $\mathbf{w}_t$. 
The turnover required to move from the drifted weights $\bar{\mathbf{w}}_t$ to $\mathbf{w}_t$ is
\begin{align}
    \mathrm{TO}_t
    =
    \sum_{i=1}^m \bigl|w_{i,t} - \bar{w}_{i,t}\bigr|,
    \label{eq:turnover_allocation}
\end{align}
incurring a proportional cost represented by
\begin{align}
    \mu_t 
    = 1 - c \cdot \mathrm{TO}_t,
    \label{eq:net_of_fee_factor}
\end{align}
where $c = 0.25\%$ is the commission rate.

\paragraph{Portfolio value update.}
The portfolio grows multiplicatively as
\begin{align}
P_t 
= P_{t-1} \cdot (\mathbf{w}_{t-1}^\top \mathbf{y}_t) \cdot \mu_t,
\label{eq:rebalanced_transaction_cost}
\end{align}
implying the one-period stochastic growth factor:
\begin{align}
    \frac{P_t}{P_{t-1}} 
    = \mu_t \cdot (\mathbf{w}_{t-1}^\top \mathbf{y}_t).
\end{align}

Taking the logarithm yields the reward in~\eqref{eq:return_reward}. 
This process repeats until the terminal period $T$, forming a trajectory 
$\tau = \{(s_t, a_t, r_t)\}_{t=1}^{T}$ used for policy optimization.

\section{Model Design}

In this paper, we utilize the innovative power of physics-informed Kolmogorov-Arnold networks (PIKANs), along with the recalculation approach of Newton's laws into physics losses, to extend standard Deep Reinforcement Learning (DRL) algorithms (DDPG, PPO, TD3, A2C). The core analogy is inspired by Newton's second law of motion ($F = m \cdot a$), adapted to a financial context for portfolio optimization. In this framework, actions ($w_t = [w_{1,t}, w_{2,t}, \dots, w_{m,t}]^T$) are interpreted as ``forces'' applied to the system. At the same time, a specific feature from the state observations (the 1-day return) serves as a proxy for ``velocity" (rate of change in asset prices/returns). The loss enforces consistency between the observed acceleration (derived from changes in velocity over time) and the predicted acceleration (force divided by mass), using mean squared error (MSE) to penalize discrepancies.
This physics-informed design encourages policies that produce actions respecting momentum-like dynamics in market returns, potentially leading to more stable and interpretable portfolios with better risk-adjusted performance (e.g., higher Sharpe ratios). The frameworks also enforce that predicted accelerations (from actions) match observed accelerations. This is done via a physics-informed loss term added to the standard RL loss, encouraging the agent to learn policies where actions align with market ``inertia" (mass $ m $) and dynamics.

\begin{sidewaysfigure}
    \centering
    \includegraphics[width=1.1\columnwidth]{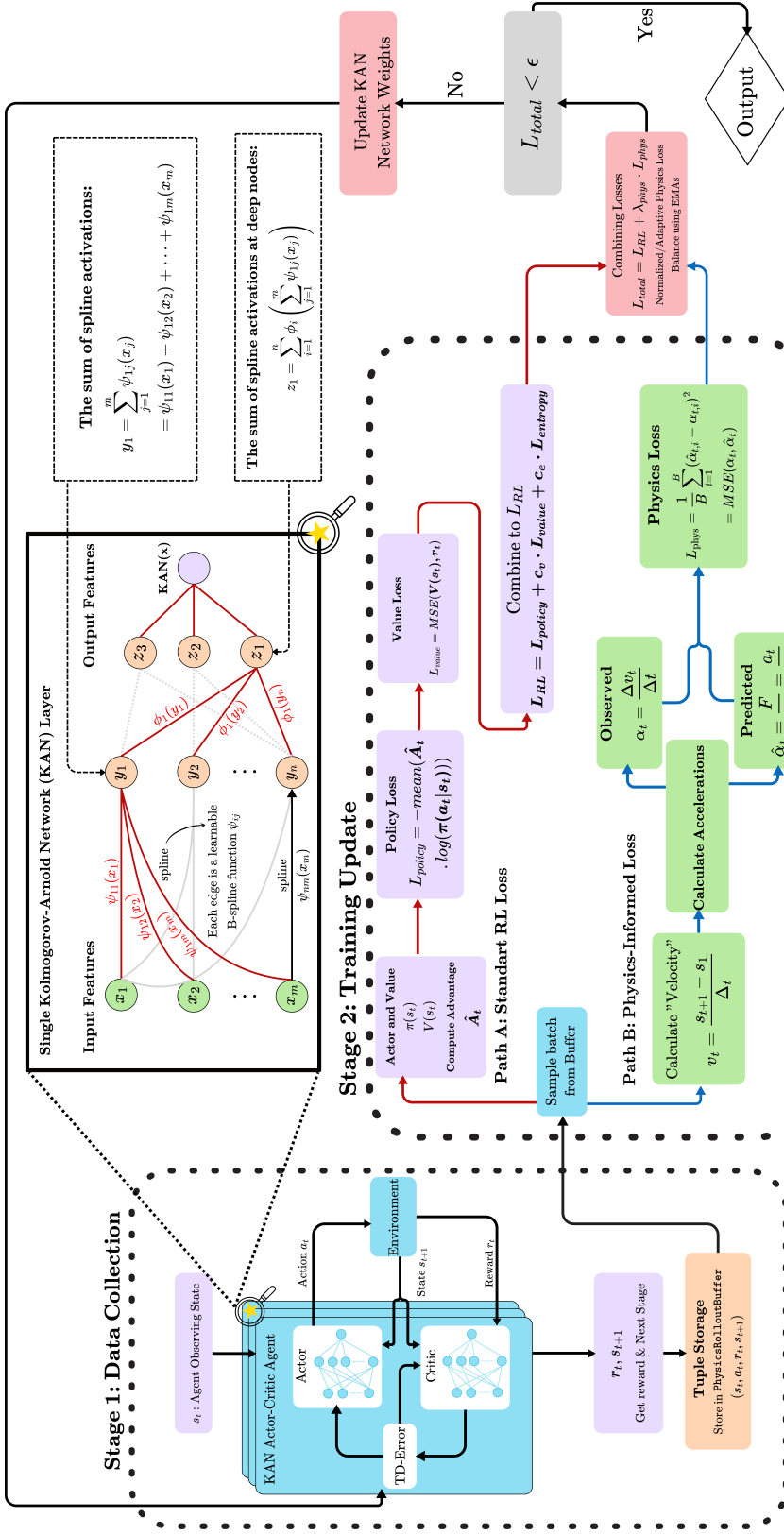}
    \caption{Overview of the physics-informed DRL framework for portfolio optimization. The environment provides states $s_t$ and price relatives $\mathbf{y}_t$, the agent outputs portfolio weights $a_t=\mathbf{w}_t$, and the portfolio value updates as $P_t=P_{t-1}(\mathbf{w}_{t-1}^\top\mathbf{y}_t)\mu_t$, 
    with transaction cost $\mu_t=1-c\,\mathrm{TO}_t$. 
    Newton’s law $F=ma$ is imposed through the physics loss $L_{\text{phys}}=\frac{1}{B}\sum_i(\hat{\alpha}_{t,i}-\alpha_{t,i})^2$, 
    where $\hat{\alpha}_t=a_t/m$ and $\alpha_t=\Delta v_t/\Delta t$.}
    \label{fig:sideways_figure}
\end{sidewaysfigure}

The general procedure for applying Newton's Laws in these frameworks begins with data collection, utilizing a custom replay or roll-out buffer to store observations ($s_t$), actions ($a_t$), rewards, and next observations ($s_{t+1}$) essential for physics computations. This is followed by policy network modification through PINN integration, where traditional Multi-Layer Perceptrons (MLPs) are replaced with Kolmogorov-Arnold Networks (KANs) that employ learnable univariate B-spline functions for enhanced parameter efficiency and interpretability, approximating functions as:

\begin{align}
    f(x) = \sum_{i=1}^{n} \phi_i \left( \sum_{j=1}^{m} \psi_{i j}(x_j) \right), \text{where }  \phi_i \text{ and } \psi_{ij} \text{ are splines}
    \label{eq:spline_approximation}
\end{align}

During training, a physics loss $L_{phys}$ is computed based on Newton's Second Law:

\begin{itemize}
    \item Compute returns (velocity analogy): $ v_t = \frac{p_{t+1} - p_t} {p_t} $,  where $ p_t $ is price from $ s_t $ (typically index 3 in features).
    \item Compute observed acceleration: $ \alpha_t = \frac{\Delta v_t}{\Delta t} $.
    \item Compute predicted acceleration: $ \hat{\alpha}_t = \frac{a_t}{m} $ (actions as forces). In the experiment, we set the asset inertia $m=1$.
    \item Physics loss: $ L_{\text{phys}} = \frac{1}{B} \sum_{i=1}^{B} (\hat{\alpha}_{t,i} - \alpha_{t,i})^2 $ (MSE), where $ B $ is batch size. In some variants, mean over features or clamp for stability. 
\end{itemize}

This physics loss is then augmented into the standard actor/policy loss via:

\begin{align}
    L_{\text{total}} = L_{\text{RL}} + \lambda_{\text{phys}} \cdot L_{\text{phys}} \label{eq:policy_loss_total}
\end{align} 

where $ \lambda_{\text{phys}} $ is a hyperparameter, which is often adaptive with normalization via Exponential Moving Averages (EMAs) for mean and variance. Training proceeds by optimizing this augmented loss, incorporating gradient clipping for stability and logging physics-related metrics such as volatility EMA as a regime proxy. Finally, evaluation demonstrates that the physics constraints enhance generalization in volatile markets by enforcing conservative, physically plausible dynamics.

Mathematical Formulations in the General Framework:

\begin{itemize}
    \item Newton's Second Law Analogy: $\hat{\alpha}_t = \frac{a_t}{m} $, where $ m $ (mass) models market inertia (e.g., resistance to price changes). Normalized/Adaptive Physics Loss: To balance scales, use Exponential Moving Averages (EMAs):
    \begin{itemize}
        \item Actor EMA:
        \begin{align}
            \mu_{\text{actor}} &=\beta \mu_{\text{actor}} + (1 - \beta) \cdot \text{mean}(Q(s_t, a_t)) \notag\\
            \sigma^2_{\text{actor}} &=\beta \sigma^2_{\text{actor}} + (1 - \beta) \cdot \text{var}(Q(s_t, a_t)) \notag
        \end{align}
        \item Physics EMA: Similar for $ L_{\text{phys}} $.
        \item Normalized: $ L_{\text{phys,norm}} = \frac{L_{\text{phys}} - \mu_{\text{phys}}}{\sqrt{\sigma^2_{\text{phys}} + \epsilon}} $.
        \item Adaptive $ \lambda $: $ \lambda_{\text{adapt}} = \lambda_{\text{phys}} \cdot \frac{|\text{mean}(L_{\text{RL}})|}{|\text{mean}(L_{\text{phys,norm}})| + \sqrt{\text{var}(L_{\text{phys,norm}})}} $, clamped for stability.
    \end{itemize}
    \item Volatility Regime Proxy: EMA volatility $ \text{Vol}_{\text{EMA}} = \beta_{\text{Vol}} \text{Vol}_{\text{EMA}} + (1 - \beta_{\text{Vol}}) \cdot \text{std}(v_t) $, penalizing high-volatility regimes.
\end{itemize}

This framework is tailored for finance, where Newton's Laws analogize market momentum and inertia.

\subsection{Deep Deterministic Policy Gradient (DDPG)}

The Deep Deterministic Policy Gradient (DDPG) algorithm, an off-policy actor-critic method for continuous action spaces, learns a deterministic policy by maximizing Q-values estimated by a critic network, with exploration via action noise and target network smoothing for stability \citep{aritonang2025hidden, sang_portfolio_2025}. In the PINN-enhanced version (\texttt{DDPG-PINN}), the structure extends DDPG-like behavior with a policy delay of 1 and no target noise, incorporating Exponential Moving Averages (EMAs) for loss normalization, adaptive weighting of the physics loss via variance penalties, and a volatility EMA proxy. Newton's Laws are applied in the actor update by computing returns as velocity, acceleration as velocity over time step, predicted acceleration as actions divided by mass, and adding the mean-squared error (MSE) physics loss (reduced per batch) to the normalized actor loss.

\begin{algorithm}[H]
\scriptsize
\caption{Adjusted DDPG with Physics-Informed Loss}
\begin{algorithmic}[1]
\For{$i = 1$ to gradient\_steps}
    \State Sample replay\_data $\sim$ ReplayBuffer(batch\_size) \Comment{$\{s_t, a_t, r_t, s_{t+1}, done\}$}
    \State // Critic Update (Standard DDPG)
    \State // Actor Update with PINN
    \State $a_{pi} \gets$ actor($s_t$)
    \State Q1 $\gets$ critic($s_t, a_{pi}$)[0]
    \State Update actor EMAs: $\mu_{actor} \gets \beta \mu_{actor} + (1-\beta)$ mean(Q1), $\sigma^2_{actor} \gets \beta \sigma^2_{actor} + (1-\beta)$ var(Q1)
    \State $L_{actor} \gets - (Q1 - \mu_{actor}) / \sqrt{\sigma^2_{actor} + \epsilon}$
    \State // Newton's Laws Physics Loss
    \State Reshape $s_t, s_{t+1}$ to [batch, window, assets, features]
    \State $v_t \gets (p_{t+1} - p_t) / p_t$ \Comment{p at feature index 3}
    \State Update vol\_ema $\gets \beta_{vol}$ vol\_ema + $(1-\beta_{vol})$ mean(std($v_t$, dim=1))
    \State $\alpha_t \gets v_t / \Delta t$
    \State $\alpha_{hat} \gets a_{pi} / m$
    \State $L_{phys raw} \gets$ MSE($\alpha_{hat}, \alpha_t$) \Comment{Clamped [0,10]}
    \State Update phys EMAs: 
    \State $\mu_{phys} \gets \beta \mu_{phys} + (1-\beta)$ mean($L_{phys raw}$), 
    \State $\sigma^2_{phys} \gets \beta \sigma^2_{phys} + (1-\beta)$ var($L_{phys raw}$)
    \State $L_{phys norm} \gets (L_{phys raw} - \mu_{phys}) / \sqrt{\sigma^2_{phys} + \epsilon}$
    \State // Adaptive Lambda
    \State mag\_actor $\gets |\text{mean}(L_{actor})|$
    \State mag\_phys $\gets |\text{mean}(L_{phys norm})|$
    \State var\_penalty $\gets \sqrt{\text{var}(L_{phys norm}) + 1e-6}$
    \State $\lambda_{adapt} \gets$ clamp( (mag\_actor / (mag\_phys + var\_penalty)) $\cdot \lambda_{phys}$, 0.1, 10 )
    \State $L_{total actor} \gets$ mean($L_{actor}$) + $\lambda_{adapt} \cdot$ mean($L_{phys norm}$)
    \State Optimize actor w.r.t. $L_{total actor}$ (clip gradients $\Vert g \Vert \leq 10$)
    \State // Polyak Update Targets
    \State Update critic\_target, actor\_target with $\tau$
\EndFor
\State Log metrics: actor\_loss, critic\_loss, physics\_loss, $\lambda_{adapt}$, vol\_ema
\end{algorithmic}
\end{algorithm}

\subsection{Twin Delayed Deep Deterministic Policy Gradient (TD3)}

The Twin Delayed Deep Deterministic Policy Gradient (TD3) extends DDPG by addressing Q-overestimation through twin critics (taking the minimum Q-value), delayed policy updates, and target policy smoothing with clipped noise \citep{aritonang2025hidden,  sang_portfolio_2025}. \texttt{TD3\_PINN} builds on the TD3's twin-critic, delayed-update framework with KAN policies, similar EMAs and adaptive lambda as \texttt{DDPG\_PINN}, but with target policy noise and clipping; Newton's Laws enforcement mirrors \texttt{DDPG\_PINN}, using per-batch MSE (mean over dimensions) for the physics loss in the actor update.

\begin{algorithm}[H]
\scriptsize
\caption{Adjusted TD3 with Physics-Informed Loss}
\begin{algorithmic}[1]
\For{$i = 1$ to gradient\_steps}
    \State Sample replay\_data $\sim$ ReplayBuffer(batch\_size) \Comment{$\{s_t, a_t, r_t, s_{t+1}, done\}$}
    \State // Critic Update (Standard Twin-DDPG)
    \If{$i \% \text{policy\_delay} == 0$}
        \State // Actor Update with PINN (Similar to DDPG\_PINN)
        \State $a_{pi} \gets$ actor($s_t$)
        \State Q1 $\gets$ critic.Q1($s_t, a_{pi}$)
        \State Update actor EMAs: $\mu_{actor}, \sigma^2_{actor}$
        \State $L_{actor} \gets - (Q1 - \mu_{actor}) / \sqrt{\sigma^2_{actor} + \epsilon}$
        \State // Newton's Laws Physics Loss (Similar to DDPG, but mean(dim=1))
        \State Reshape $s_t, s_{t+1}$; $v_t \gets (p_{t+1} - p_t)/(p_t + \epsilon)$
        \State Update vol\_ema
        \State $\alpha_t \gets v_t / \Delta t$
        \State $\alpha_{hat} \gets a_{pi} / m$
        \State $L_{phys raw} \gets$ mean(MSE($\alpha_{hat}, \alpha_t$, reduction='none'), dim=1) \Comment{Per batch, clamped}
        \State Update phys EMAs: $\mu_{phys}, \sigma^2_{phys}$ (using mean/var over batch)
        \State $L_{phys norm} \gets (L_{phys raw} - \mu_{phys}) / \sqrt{\sigma^2_{phys} + \epsilon}$
        \State // Adaptive Lambda (Same as DDPG)
        \State $\lambda_{adapt} \gets ...$
        \State $L_{total actor} \gets$ mean($L_{actor}$) + $\lambda_{adapt} \cdot$ mean($L_{phys norm}$)
        \State Optimize actor w.r.t. $L_{total actor}$ (clip $\Vert g \Vert \leq 10$)
        \State Polyak update actor\_target with $\tau$
    \EndIf
\EndFor
\State Log metrics: Similar to \texttt{DDPG\_PINN}
\end{algorithmic}
\end{algorithm}

\subsection{Proximal Policy Optimization (PPO)}

The Proximal Policy Optimization (PPO) algorithm, an on-policy method, optimizes a clipped surrogate objective to ensure small policy updates, balancing exploration and exploitation while estimating advantages with generalized advantage estimation (GAE) \citep{aritonang2025hidden, sang_portfolio_2025}. \texttt{PPO\_PINN} modifies the on-policy structure with a custom rollout buffer to store next observations and a KAN-based policy, fixing the physics weight without adaptation; it incorporates Newton's Laws by calculating velocity from state differences over time step, averaging last-step velocity over features for acceleration, and using MSE between actions (as acceleration, implying mass=1) and observed values in the total policy loss.

\begin{algorithm}[H]
\scriptsize
\caption{Adjusted PPO with Physics-Informed Loss}
\begin{algorithmic}[1]
\For{epoch = 1 to n\_epochs}
    \For{rollout\_data in PhysicsRolloutBuffer.get(batch\_size)} \Comment{$\{s_t, a_t, old_V, old_logp, adv, returns, s_{t+1}\}$}
        \State // Physics Loss (Newton's Laws)
        \State Reshape $s_t, s_{t+1}$ to [batch, window, assets, features]
        \State $v \gets (s_{t+1} - s_t) / \Delta t$
        \State last\_v $\gets v[:, -1, :, :]$ \Comment{Last time step}
        \State $\alpha \gets$ mean(last\_v, dim=2) \Comment{Mean over features}
        \State $L_{phys} \gets$ MSE($a_t, \alpha$) \Comment{Actions as accel (m=1)}
        \State // Policy Evaluation
        \State V, logp, entropy $\gets$ policy.evaluate\_actions($s_t, a_t$)
        \If{Box action} logp $\gets$ sum(logp, dim=1) \EndIf
        \State adv\_norm $\gets (adv - \text{mean}(adv)) / (\text{std}(adv) + \epsilon)$ if normalize else adv
        \State ratio $\gets \exp(\text{logp} - old_logp)$
        \State $L_{policy1} \gets$ adv\_norm $\cdot$ ratio
        \State $L_{policy2} \gets$ adv\_norm $\cdot$ clamp(ratio, $1-\epsilon_{clip}, 1+\epsilon_{clip}$)
        \State $L_{policy} \gets -$mean(min($L_{policy1}, L_{policy2}$))
        \State // Value Loss
        \State V\_clip $\gets old_V +$ clamp(V $- old_V, -\epsilon_{vf}, +\epsilon_{vf}$) if clip\_vf else V
        \State $L_{value} \gets$ MSE(V\_clip, returns)
        \State // Entropy Loss
        \State $L_{entropy} \gets -$mean(entropy) if entropy else $-$mean($-$logp)
        \State $L_{total} \gets L_{policy} +$ ent\_coef $\cdot L_{entropy} +$ vf\_coef $\cdot L_{value} + \lambda_{phys} \cdot L_{phys}$
        \State Optimize policy w.r.t. $L_{total}$ (clip gradients $\Vert g \Vert \leq$ max\_grad\_norm)
        \State // Early Stop if KL $> target_kl \cdot 1.5$
    \EndFor
\EndFor
\State Log metrics: policy\_loss, value\_loss, entropy\_loss, physics\_loss, explained\_var
\end{algorithmic}
\end{algorithm}

\subsection{Advantage Actor-Critic (A2C)}

The Advantage Actor-Critic (A2C) algorithm, a synchronous variant of A3C, uses parallel environments to compute advantages and update an actor-critic policy on-policy, with entropy regularization for exploration \citep{qiu2024design}. \texttt{A2C\_PINN} enhances the synchronous actor-critic with a custom buffer and dynamic scaling of physics loss based on policy loss magnitude, normalizing observations and velocities for stability; it applies Newton's Laws by deriving acceleration from averaged feature velocities over time step, predicting it as actions over mass, and adding MSE plus variance to the total loss for full-buffer updates.

\begin{algorithm}[H]
\scriptsize
\caption{Adjusted A2C with Physics-Informed Loss}
\begin{algorithmic}[1]
\For{rollout\_data in PhysicsRolloutBuffer.get()} \Comment{Full buffer}
    \State // Physics Loss (Newton's Laws)
    \State Reshape $s_t, s_{t+1}$ to [batch, window, assets, features]
    \State Normalize: $s \gets (s - \text{mean}(s)) / (\text{std}(s) + \epsilon)$
    \State $v \gets (s_{t+1} - s_t) / \Delta t$; Normalize v
    \State $\alpha \gets$ mean($v[:, -1, :, :] / \Delta t$, dim=2) \Comment{Mean over features}
    \State $\alpha_{hat} \gets a_t.unsqueeze(2) / m$
    \State $L_{phys} \gets$ MSE($\alpha_{hat}, \alpha.unsqueeze(2)$) + var($L_{phys}$) \Comment{Clamped [-10,10]}
    \State // Policy Evaluation
    \State V, logp, entropy $\gets$ policy.evaluate\_actions($s_t, a_t$)
    \If{Box} logp $\gets$ sum(logp, dim=1); entropy $\gets$ sum(entropy, dim=1) if entropy \EndIf
    \State adv\_norm $\gets (adv - \text{mean}(adv)) / (\text{std}(adv) + \epsilon)$ if normalize
    \State $L_{policy} \gets -$mean(adv\_norm $\cdot$ logp)
    \State $L_{value} \gets$ MSE(V.flatten(), returns)
    \State $L_{entropy} \gets -$mean(entropy) if entropy else $-$mean($-$logp)
    \State // Dynamic Scaling
    \State mag\_policy $\gets |L_{policy}|$
    \State mag\_phys $\gets |L_{phys}|$
    \State scale $\gets$ mag\_policy / (mag\_phys + $\epsilon$)
    \State $L_{total} \gets L_{policy} +$ ent\_coef $\cdot L_{entropy} +$ vf\_coef $\cdot L_{value} + \lambda_{phys} \cdot$ scale $\cdot L_{phys}$
    \State Optimize policy w.r.t. $L_{total}$ (clip $\Vert g \Vert \leq$ max\_grad\_norm)
\EndFor
\State Log metrics: policy\_loss, value\_loss, entropy\_loss, physics\_loss, explained\_var
\end{algorithmic}
\end{algorithm}

\section{Experimental setting}

\label{sec:experimental_setting}
\subsection{Dataset}
The model is trained and evaluated using historical daily stock data from three distinct international market universes: the China market (CSI 300), Vietnam market (VN100), and U.S market (S\&P 100). This selection provides a comprehensive cross-section of global equity markets, encompassing diverse volatility profiles and economic growth stages. 

As detailed in Table \ref{tab:portfolios}, each portfolio consists of 10 liquid assets. To ensure a consistent comparison, all portfolios share a common training period from 2015-01-01 to 2023-01-01 and an out-of-sample testing period from 2023-01-02 to 2025-01-01.

\begin{table}[h!]
\centering
\caption{Portfolio Composition and Experimental Timeframes}
\label{tab:portfolios}
\small
\begin{tabularx}{\textwidth}{@{}l X@{}}
\toprule
\textbf{Portfolio} & \textbf{Asset Tickers} \\ \midrule
China market & 601868.SS, 600010.SS, 601669.SS, 300059.SZ, 601398.SS, 600111.SS, 603993.SS, 600050.SS, 600515.SS, 600875.SS \\ \addlinespace
Vietnam market & HPG.VN, VIX.VN, MWG.VN, DIG.VN, MSN.VN, STB.VN, VSC.VN, HSG.VN, FPT.VN, EIB.VN \\ \addlinespace
U.S. market & MSFT, INTC, BAC, DIS, PFE, LLY, NKE, VZ, WMT, AMGN \\ \midrule
\multicolumn{2}{@{}l@{}}{\textbf{Experimental Periods:}} \\
Training Period & 2015-01-01 -- 2023-01-01 \\
Testing Period & 2023-01-02 -- 2025-01-01 \\ \bottomrule
\end{tabularx}
\end{table}

\subsection{Performance metrics}

To evaluate the effectiveness of the proposed portfolio allocation strategies, we adopt a multi-dimensional set of quantitative indicators. These metrics assess performance across total profitability, risk-adjusted returns, and downside protection, following established benchmarks in financial deep reinforcement learning \cite{jiang_cryptocurrency_2016}. The primary measures—Cumulative Return Rate (CRR), Annualized Return (AR), Annualized Volatility (AV), Sharpe Ratio (SR), Maximum Drawdown (MDD), and Calmar Ratio (CR)—are defined below.

\paragraph{Cumulative Return Rate (CRR)}
The CRR quantifies the total percentage growth of the portfolio wealth over the entire testing horizon:
\begin{equation}
    CRR = \frac{P_{T} - P_{0}}{P_{0}} \times 100\%
\end{equation}
where $P_{0}$ and $P_{T}$ represent the initial and final portfolio values, respectively.

\paragraph{Annualized Return (AR)}
The annualized return, also called the geometric average return, measures the compound rate of return on a yearly basis:
\begin{align}
    Annualized\ Return = \left( \frac{p_{t_f}}{p_0} \right)^{\frac{1}{years}} - 1
    \label{eq:annual_return}
\end{align}
where $years$ is the number of years in the testing period.

\paragraph{Annualized Volatility (AV)}
Annualized volatility quantifies the annualized standard deviation of the portfolio’s daily returns. 
If $\sigma_{daily}$ denotes the standard deviation of daily returns, and $N_{trading}=252$ represents the average number of trading days in a year, then:
\begin{align}
    \sigma_{annual} = \sigma_{daily} \times \sqrt{N_{trading}}
    \label{eq:annual_volatility}
\end{align}

\paragraph{Sharpe Ratio (SR)}
Proposed by \cite{sharpe_sharpe_1994}, the Sharpe ratio evaluates the risk-adjusted performance of the portfolio and is computed as:
\begin{align}
    Sharpe = \frac{R_p - R_f}{\sigma_p}
    \label{eq:sharpe}
\end{align}
where $R_p$ is the average portfolio return, $R_f$ is the risk-free rate (assumed to be zero in this study), and $\sigma_p$ is the portfolio's annualized volatility.

\paragraph{Maximum Drawdown (MDD)}
The maximum drawdown measures the largest peak-to-trough decline in the portfolio value during the test period \citep{magdon-ismail_malik_and_atiya_amir_f_maximum_2004}. It is expressed as:
\begin{align}
    MDD = \max_{t \in [0, T]} \left( \frac{p_{peak} - p_t}{p_{peak}} \right) \times 100(\%)
    \label{eq:maximum_drawdown}
\end{align}

\paragraph{Calmar Ratio (CR)} 
The Calmar ratio \cite{terry_w_calmar_1991} evaluates the trade-off between return and drawdown risk. 
It is defined as the ratio between annualized return and the absolute value of the maximum drawdown:
\begin{align}
    Calmar = \frac{Annualized\ Return}{|MDD|}
    \label{eq:calmar}
\end{align}

All these indicators together provide a comprehensive evaluation of both profitability and risk for portfolio strategies trained using DRL.
\section{Results and Discussion}
\label{sec:results}


\subsection{Quantitative Performance Comparison Across Markets}

Tables~\ref{tab:perf_china}--\ref{tab:perf_us} summarize the performance of all methods across the China, Vietnam and U.S. markets. The reported metrics include cumulative return (CumRet), annualized return (AnnRet), annualized volatility (AnnVol), Sharpe ratio, maximum drawdown (MDD), Calmar ratio, and average portfolio value (APV).

Across all three markets, the physics-informed PIKAN agents either outperform or match their vanilla DRL counterparts and classical online portfolio-selection baselines. In the U.S.\ market (Table~\ref{tab:perf_us}), \texttt{A2C\_PINN} attains a cumulative return of 44.66\%, an annualized return of 20.59\%, a Sharpe ratio of 1.54, and a Calmar ratio of 1.73. These values clearly exceed those of standard \texttt{A2C} (CumRet 10.42\%, Sharpe 0.36, Calmar 0.31) and traditional strategies such as \texttt{CRP} (CumRet 16.22\%, Sharpe 0.60) and \texttt{UBAH} (CumRet 22.23\%, Sharpe 0.81). The physics constraint thus improves both profitability and risk-adjusted performance in a relatively mature market.

In the Vietnamese market (Table~\ref{tab:perf_vn}), \texttt{TD3\_PINN} achieves the best overall performance with a 37.36\% cumulative return, 17.58\% annualized return, Sharpe ratio of 0.68, and Calmar ratio of 0.67. Its vanilla counterpart, \texttt{TD3}, yields a negative cumulative return of --11.30\% and a Sharpe ratio of --0.28, highlighting the role of the Newtonian regularizer in stabilizing learning under higher volatility and weaker market efficiency. \texttt{DDPG\_PINN} and \texttt{SAC} also show strong positive returns and Sharpe ratios in this market, further supporting the benefit of physics-informed losses for off-policy actor--critic methods.

The CSI~300 results (Table~\ref{tab:perf_china}) illustrate the resilience of the proposed approach under challenging conditions. Several vanilla DRL agents (e.g., \texttt{PPO}, \texttt{TD3}, \texttt{DDPG}) and adversarial online strategies (e.g., \texttt{OLMAR}, \texttt{RMR}) incur substantial losses. In contrast, \texttt{A2C\_PINN} maintains a positive cumulative return of 17.29\% with a Sharpe ratio of 0.44 and a Calmar ratio of 0.41, outperforming not only its vanilla counterpart (\texttt{A2C}, CumRet --12.65\%) but also classical benchmarks such as \texttt{CRP} and \texttt{UBAH}. This pattern suggests that the inertia-like constraint helps regularize the policy in non-stationary and noisy environments where purely data-driven models tend to overfit short-term fluctuations.

Taken together, these results indicate that the Newtonian physics-informed term acts as a useful inductive bias: it suppresses extreme reallocations, moderates risk exposure, and improves risk-adjusted efficiency across different levels of market development.
\begin{table}[t]
\centering
\caption{Performance metrics across algorithms in the Chinese market (CSI~300).}
\label{tab:perf_china}
\begin{tabular}{lrrrrrrr}
\toprule
Algorithm & CumRet (\%) & AnnRet (\%) & AnnVol (\%) & Sharpe & MDD (\%) & Calmar & APV \\
\midrule
PPO\_PINN & -34.72 & -20.10 & 21.18 & -0.95 & -48.05 & -0.42 & 652.82 \\
TD3\_PINN & -16.52 & -9.06 & 20.58 & -0.44 & -31.42 & -0.29 & 834.79 \\
DDPG\_PINN & 5.60 & 2.91 & 19.84 & 0.15 & -26.74 & 0.11 & 1056.00 \\
\textbf{A2C\_PINN} & \textbf{17.29} & \textbf{8.75} & \textbf{20.02} & \textbf{0.44} & \textbf{-21.12} & \textbf{0.41} & \textbf{1172.88} \\
PPO & -56.14 & -35.18 & 19.54 & -1.80 & -61.11 & -0.58 & 438.65 \\
TD3 & -39.16 & -23.00 & 20.90 & -1.10 & -49.05 & -0.47 & 608.45 \\
DDPG & -39.16 & -23.00 & 20.90 & -1.10 & -49.05 & -0.47 & 608.45 \\
SAC & -33.91 & -19.58 & 23.58 & -0.83 & -50.71 & -0.39 & 660.89 \\
A2C & -12.65 & -6.87 & 21.76 & -0.32 & -37.24 & -0.18 & 873.46 \\
UBAH & 7.50 & 3.85 & 19.52 & 0.20 & -26.51 & 0.15 & 1075.01 \\
CRP & 7.52 & 3.85 & 19.78 & 0.19 & -26.74 & 0.14 & 1075.17 \\
OLMAR & -74.92 & -51.40 & 30.94 & -1.66 & -78.31 & -0.66 & 250.81 \\
RMR & -74.13 & -50.61 & 30.87 & -1.64 & -76.79 & -0.66 & 258.74 \\
PAMR & 7.52 & 3.85 & 19.78 & 0.19 & -26.74 & 0.14 & 1075.17 \\
\bottomrule
\end{tabular}
\end{table}
\begin{figure}[H]
  \centering
  \includegraphics[width=\textwidth]{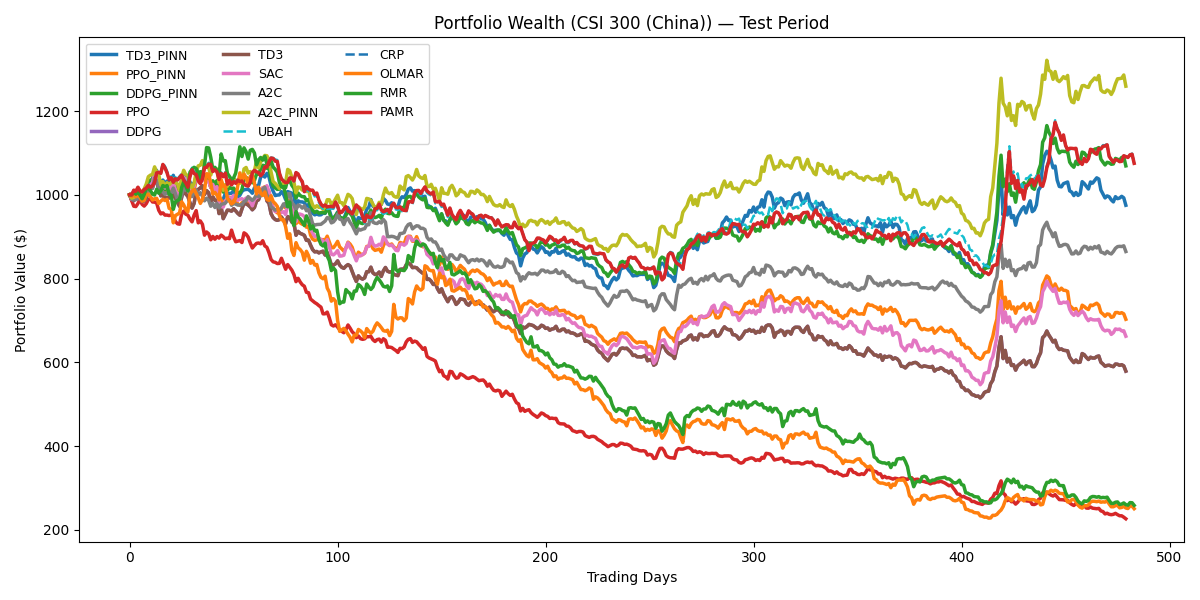}
  \caption{Portfolio wealth trajectories for CSI~300 (China) during the test period. The physics-informed variants show stronger resilience in volatile market conditions.}
  \label{fig:wealth_china}
\end{figure}

\begin{table}[t]
\centering
\caption{Performance metrics across algorithms in the Vietnam market (VN100).}
\label{tab:perf_vn}
\begin{tabular}{lrrrrrrr}
\toprule
Algorithm & CumRet (\%) & AnnRet (\%) & AnnVol (\%) & Sharpe & MDD (\%) & Calmar & APV \\
\midrule
PPO\_PINN & -39.73 & -22.77 & 23.07 & -0.99 & -43.92 & -0.52 & 602.66 \\
\textbf{TD3\_PINN} & \textbf{37.36} & \textbf{17.58} & \textbf{25.70} & \textbf{0.68} & \textbf{-26.25} & \textbf{0.67} & \textbf{1373.62} \\
DDPG\_PINN & 29.86 & 14.26 & 22.85 & 0.62 & -25.79 & 0.55 & 1298.55 \\
A2C\_PINN & 8.79 & 4.39 & 29.94 & 0.15 & -38.71 & 0.11 & 1087.87 \\
PPO & -52.72 & -31.76 & 26.20 & -1.21 & -59.70 & -0.53 & 472.78 \\
TD3 & -11.30 & -5.94 & 21.51 & -0.28 & -33.84 & -0.18 & 886.95 \\
DDPG & -11.30 & -5.94 & 21.51 & -0.28 & -33.84 & -0.18 & 886.95 \\
SAC & 33.60 & 15.93 & 24.38 & 0.65 & -27.47 & 0.58 & 1336.04 \\
A2C & 3.08 & 1.56 & 22.12 & 0.07 & -27.70 & 0.06 & 1030.77 \\
UBAH & 33.48 & 15.73 & 24.73 & 0.64 & -28.12 & 0.56 & 1334.76 \\
CRP & 29.46 & 13.96 & 22.80 & 0.61 & -25.79 & 0.54 & 1294.61 \\
OLMAR & -73.04 & -48.49 & 41.29 & -1.17 & -73.95 & -0.66 & 269.59 \\
RMR & -70.96 & -46.51 & 41.33 & -1.13 & -74.55 & -0.62 & 290.38 \\
PAMR & 29.46 & 13.96 & 22.80 & 0.61 & -25.79 & 0.54 & 1294.61 \\
\bottomrule
\end{tabular}
\end{table}
\begin{figure}[H]
  \centering
  \includegraphics[width=\textwidth]{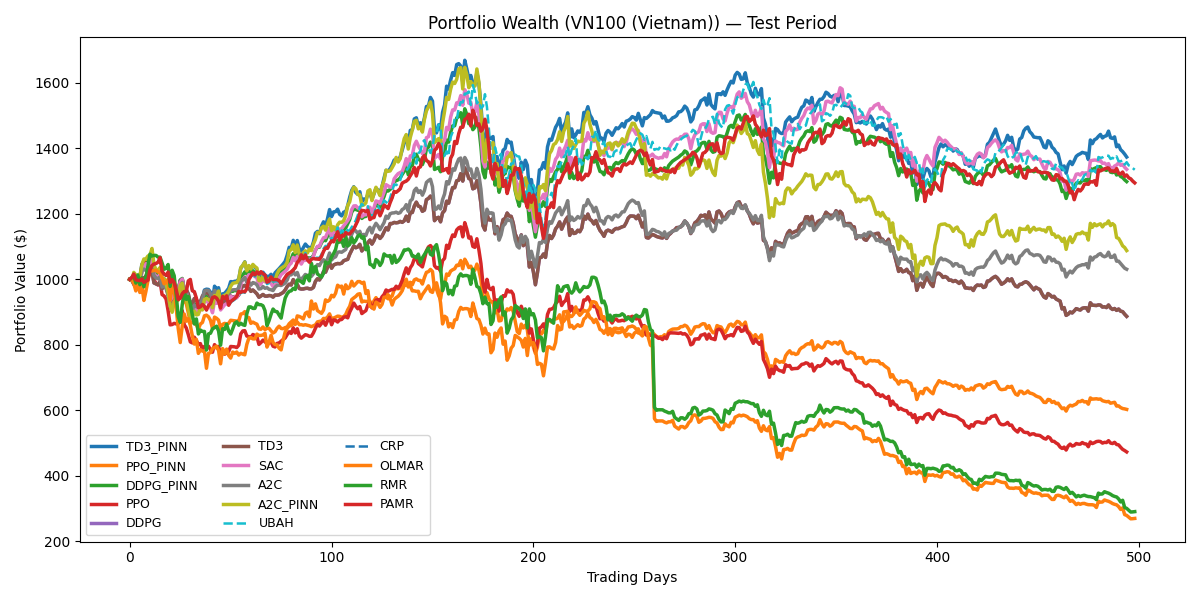}
  \caption{Portfolio wealth trajectories for Vietnam market during the test period. Physics-informed DRL agents maintain higher stability and superior cumulative returns compared with non-physics models.}
  \label{fig:wealth_vn}
\end{figure}

\begin{table}[t]
\centering
\caption{Performance metrics across algorithms in the U.S. market (S\&P~100).}
\label{tab:perf_us}
\begin{tabular}{lrrrrrrr}
\toprule
Algorithm & CumRet (\%) & AnnRet (\%) & AnnVol (\%) & Sharpe & MDD (\%) & Calmar & APV \\
\midrule
PPO\_PINN & -45.14 & -26.25 & 19.66 & -1.34 & -46.41 & -0.57 & 548.57 \\
TD3\_PINN & 25.65 & 12.27 & 12.11 & 1.01 & -9.96 & 1.23 & 1256.49 \\
DDPG\_PINN & 14.61 & 7.16 & 12.91 & 0.55 & -12.43 & 0.58 & 1146.07 \\
\textbf{A2C\_PINN} & \textbf{44.66} & \textbf{20.59} & \textbf{13.35} & \textbf{1.54} & \textbf{-11.89} & \textbf{1.73} & \textbf{1446.58} \\
PPO & -57.29 & -35.04 & 15.95 & -2.20 & -58.22 & -0.60 & 427.12 \\
TD3 & -8.55 & -4.43 & 12.60 & -0.35 & -13.77 & -0.32 & 914.50 \\
DDPG & -8.55 & -4.43 & 12.60 & -0.35 & -13.77 & -0.32 & 914.50 \\
SAC & 8.21 & 4.08 & 17.26 & 0.24 & -19.77 & 0.21 & 1082.11 \\
A2C & 10.42 & 5.16 & 14.43 & 0.36 & -16.80 & 0.31 & 1104.25 \\
UBAH & 22.23 & 10.62 & 13.18 & 0.81 & -13.59 & 0.78 & 1222.29 \\
CRP & 16.22 & 7.85 & 12.99 & 0.60 & -12.43 & 0.63 & 1162.22 \\
OLMAR & -85.41 & -62.02 & 35.21 & -1.76 & -85.83 & -0.72 & 145.89 \\
RMR & -83.47 & -59.57 & 35.15 & -1.69 & -83.89 & -0.71 & 165.27 \\
PAMR & 16.22 & 7.85 & 12.99 & 0.60 & -12.43 & 0.63 & 1162.22 \\
\bottomrule
\end{tabular}
\end{table}
\begin{figure}[H]
  \centering
  \includegraphics[width=\textwidth]{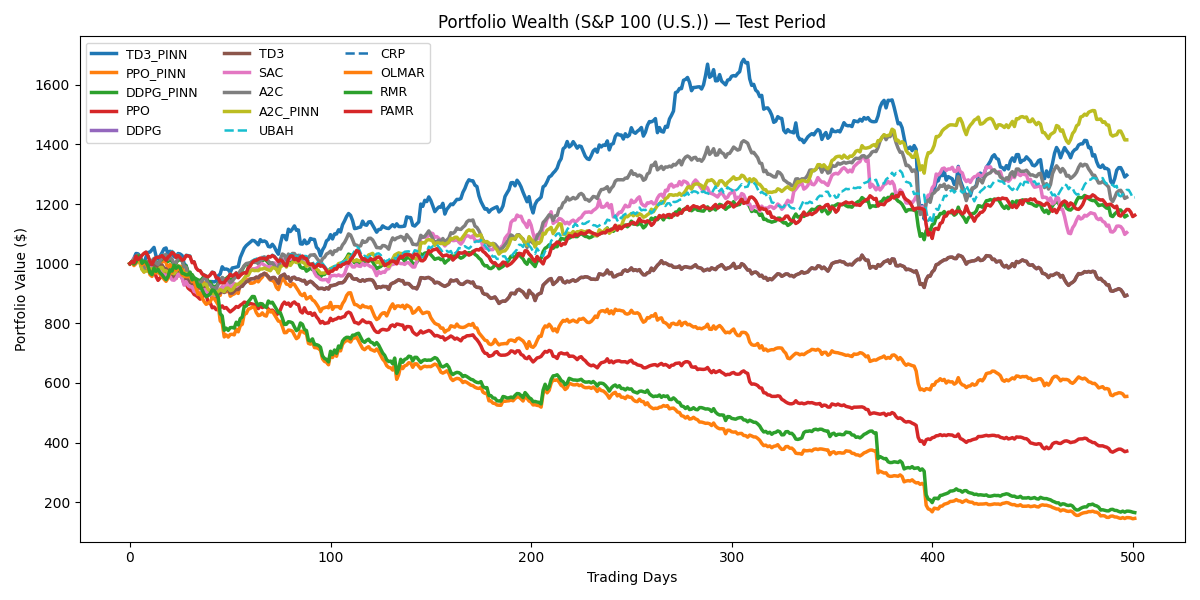}
  \caption{Portfolio wealth trajectories for S\&P~100 (U.S.) during the test period. \texttt{A2C\_PINN} and \texttt{TD3\_PINN} outperform other DRL and rule-based methods in both return and risk control.}
  \label{fig:wealth_us}
\end{figure}
\subsection{Portfolio Wealth Trajectories}

Figures~\ref{fig:wealth_vn}--\ref{fig:wealth_china} plot the evolution of portfolio wealth for all methods on China, Vietnam and U.S. markets, starting from an initial capital of \$1{,}000. In each market, the physics-informed agents produce smoother equity curves and avoid the deep drawdowns exhibited by some vanilla DRL and rule-based strategies.

In the U.S.\ market (Figure~\ref{fig:wealth_us}), \texttt{A2C\_PINN} and \texttt{TD3\_PINN} steadily compound wealth over time, with much shallower drawdowns than \texttt{PPO}, \texttt{TD3}, and \texttt{DDPG}. The physics-informed curves track above \texttt{CRP}, \texttt{UBAH}, and other online portfolio-selection methods for most of the test horizon, consistent with the higher Sharpe and Calmar ratios reported in Table~\ref{tab:perf_us}.

A similar pattern is observed in Vietnam (Figure~\ref{fig:wealth_vn}): \texttt{TD3\_PINN} and \texttt{DDPG\_PINN} grow wealth more smoothly than their vanilla counterparts and exhibit smaller peak-to-trough declines than adversarial strategies such as \texttt{OLMAR} and \texttt{RMR}, which suffer large losses in turbulent periods. In China (Figure~\ref{fig:wealth_china}), where many DRL agents experience long drawdowns, the physics-informed variants maintain relatively stable trajectories and avoid catastrophic capital erosion.

The profit--loss ratio comparison in Figure~\ref{fig:pl_ratio_us} reinforces these observations: physics-informed variants deliver more favorable profit--loss profiles, indicating that improvements are not driven by a small number of extreme gains but by systematically better trade-offs between gains and losses.

\subsection{Portfolio-Weight Dynamics and Learning Behavior}

\begin{figure}[H]
  \centering
  \includegraphics[width=0.7\textwidth]{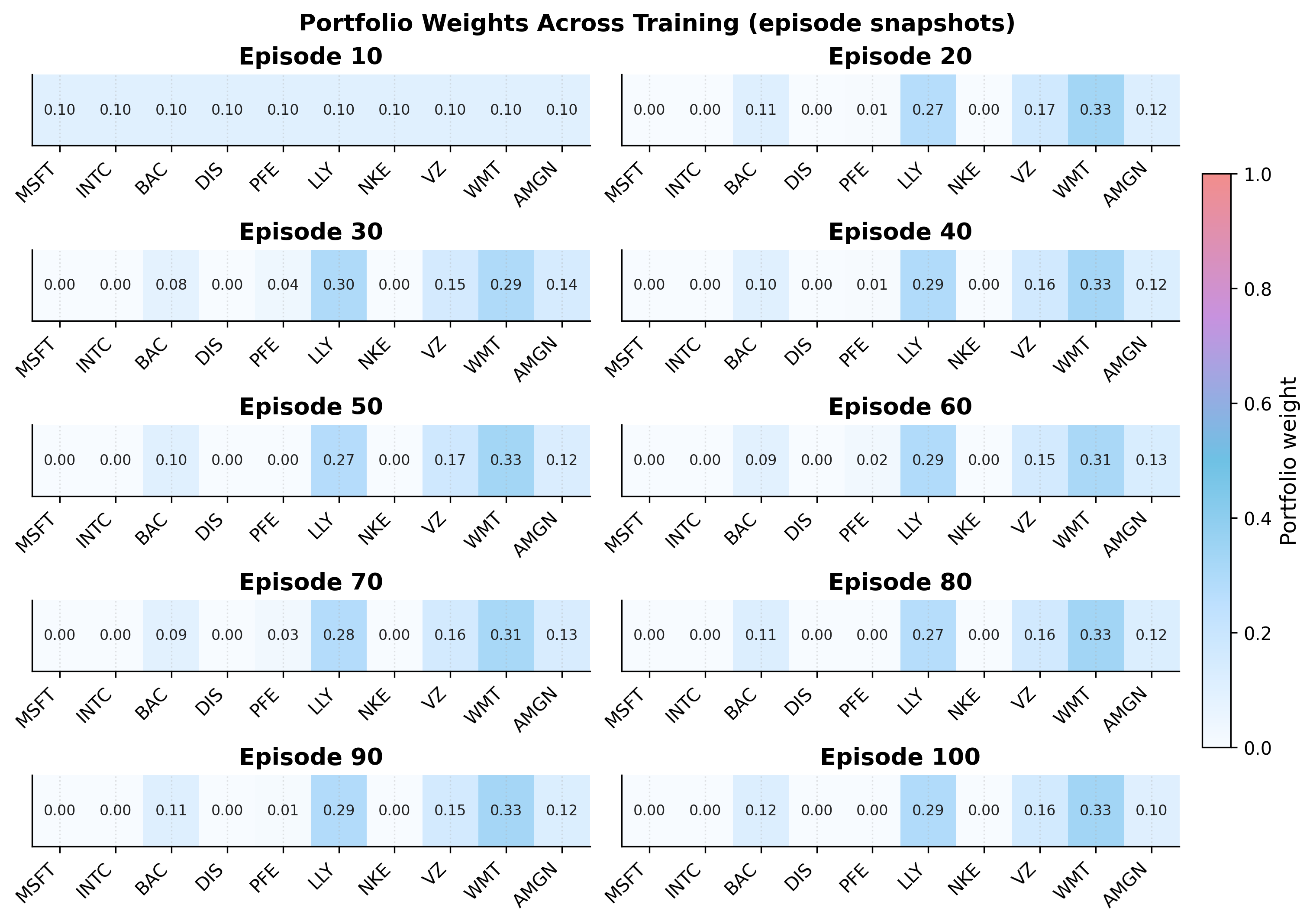}
  \caption{Evolution of portfolio weights across training episodes for \texttt{A2C\_PINN} (U.S. market).}
  \label{fig:weights_pinn}
\end{figure}

\begin{figure}[H]
  \centering
  \includegraphics[width=0.7\textwidth]{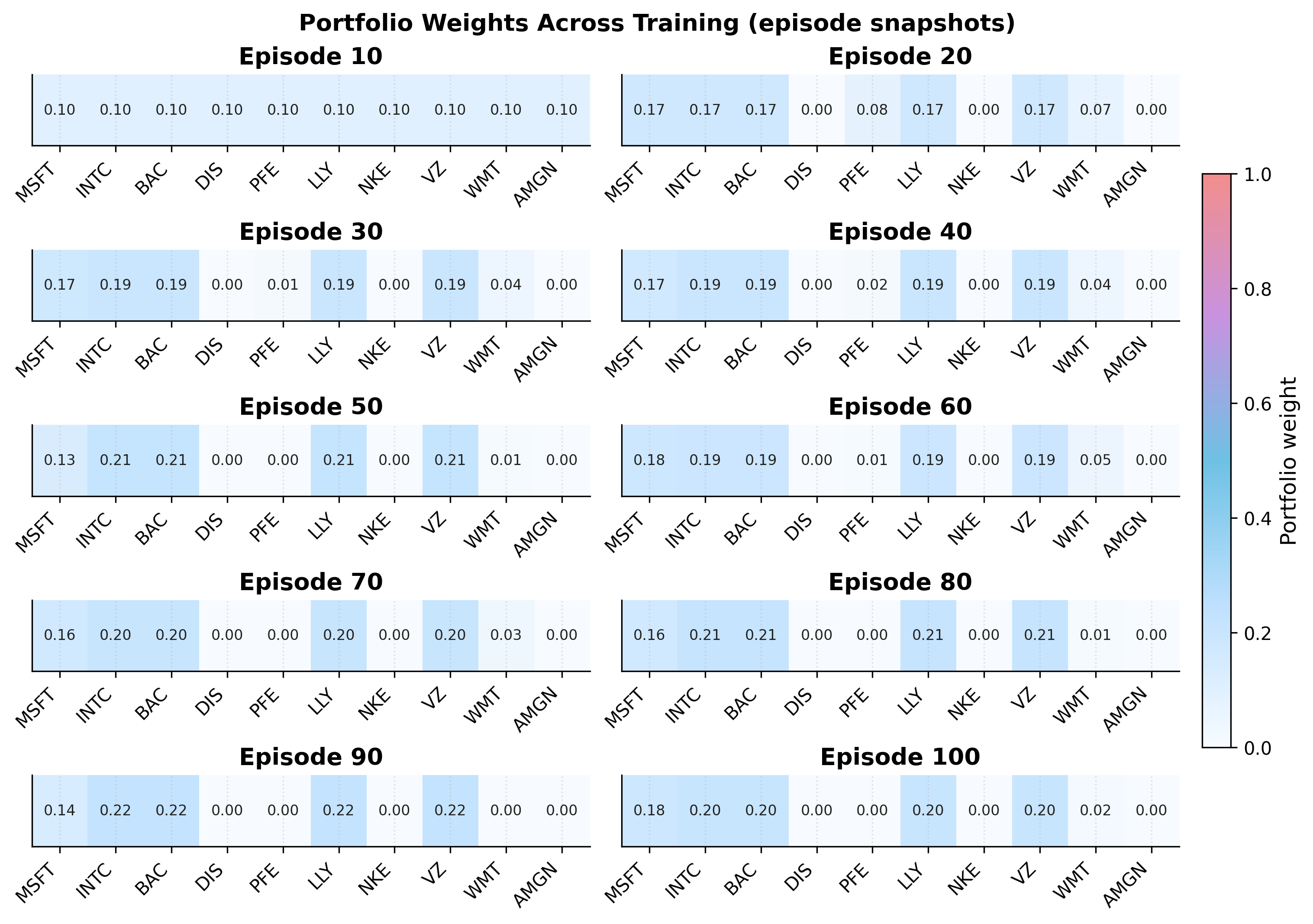}
  \caption{Evolution of portfolio weights across training episodes for \texttt{A2C} (U.S. market).}
  \label{fig:weights_a2c}
\end{figure}

Figures~\ref{fig:weights_pinn} and~\ref{fig:weights_a2c} provide further insight into the learning behavior by visualizing the evolution of portfolio weights over training episodes for \texttt{A2C\_PINN} and vanilla \texttt{A2C} on the U.S.\ dataset. Each heatmap snapshot shows the allocation to ten representative assets at different points in training.

The physics-informed \texttt{A2C\_PINN} quickly converges toward stable allocation patterns. By intermediate episodes, the agent assigns higher weights to assets exhibiting persistent trends and reduces exposure to noisy series. Over later episodes, weight adjustments become incremental rather than abrupt, reflecting a policy that respects inertia-like dynamics in returns. This temporal coherence is consistent with the Newtonian design: by penalizing discrepancies between observed accelerations (changes in returns) and action-implied accelerations, the physics loss discourages sudden, large portfolio shifts unless strongly supported by the data.

\begin{figure}[H]
  \centering
  \includegraphics[width=0.7\textwidth]{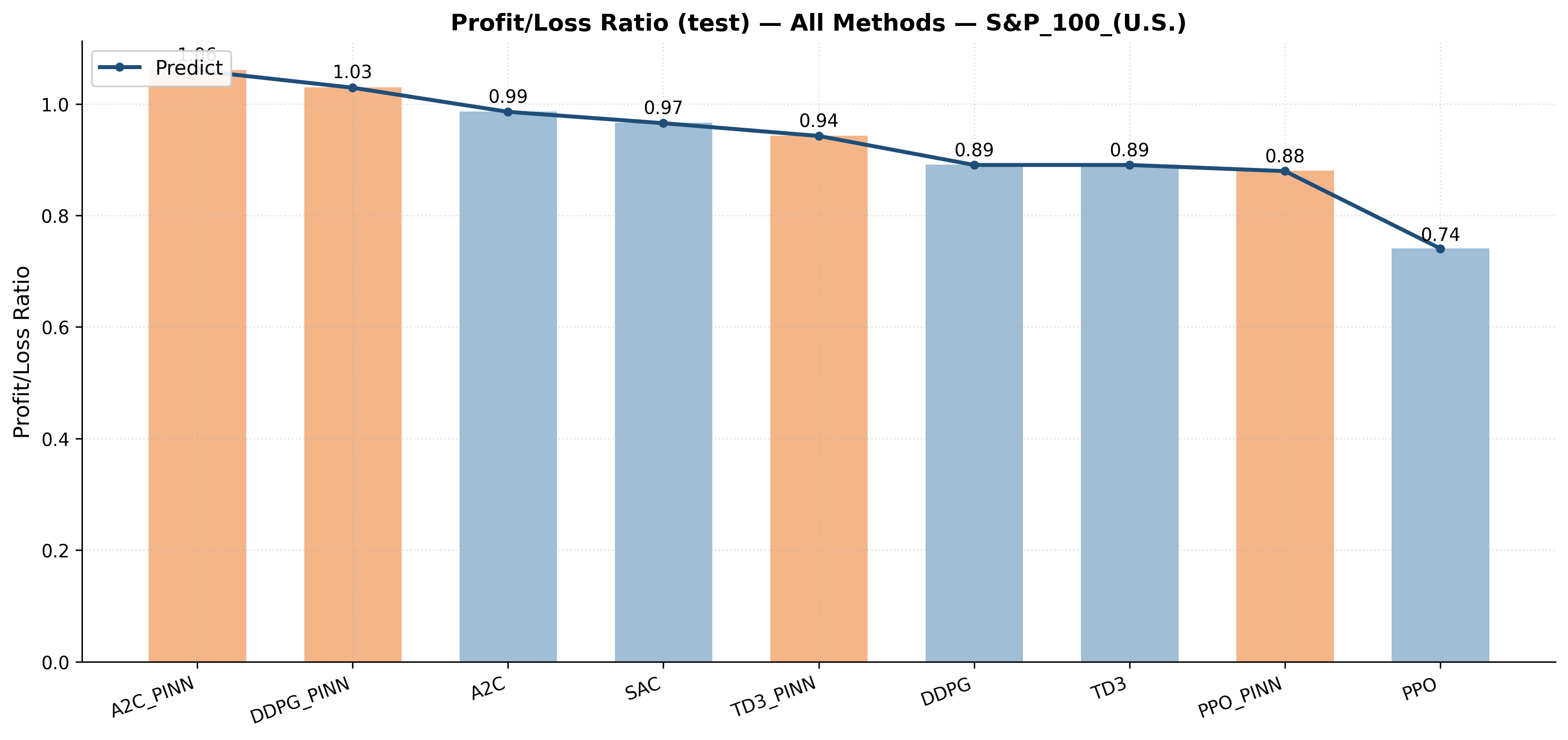}
  \caption{Profit–loss ratio comparison across all methods on the S\&P~100 (U.S.) test set. Physics-informed variants achieve superior risk-adjusted efficiency.}
  \label{fig:pl_ratio_us}
\end{figure}

In contrast, the vanilla \texttt{A2C} agent displays pronounced fluctuations in allocation over episodes. Weights frequently oscillate between assets with little persistence, mirroring the noisier and less robust wealth trajectories observed in Figures~\ref{fig:wealth_us} and~\ref{fig:pl_ratio_us}. These patterns suggest that the absence of a physics prior allows the policy to overreact to short-term signals, especially under transaction costs and non-stationary price dynamics.

\subsection{Interpretation and Limitations}

The empirical results show that the Newtonian physics-informed loss operates as an effective domain-specific regularizer for financial RL. By treating returns as velocities, actions as forces, and a tunable mass as market inertia, the framework promotes smoother and more coherent adjustments in portfolio weights. This aligns with broader evidence in physics-informed learning, where embedding simplified governing laws improves robustness and generalization \citep{raissi_physics-informed_2019, karniadakis_physics-informed_2021, cuomo_scientific_2022, banerjee_survey_2025}.

The Newtonian abstraction remains simplified and does not capture liquidity, microstructure effects, or nonlinear market regimes, and the PIKAN architecture introduces hyperparameters that require calibration. Nonetheless, the findings indicate that even lightweight physical priors can enhance both risk-adjusted performance and interpretability, offering a promising direction for more stable and principled RL-based portfolio strategies.

\section{Conclusion}
\label{sec:conclusion}
This paper proposed a physics-informed Kolmogorov--Arnold Network (PIKAN) framework for portfolio management, in which a Newtonian loss and spline-based architectures are integrated into standard actor--critic reinforcement learning. Across three markets, the physics-informed variants consistently improved cumulative returns, Sharpe ratios, and drawdown profiles relative to both vanilla DRL agents and classical portfolio-selection baselines. The learned policies also exhibited smoother and more coherent portfolio-weight dynamics, indicating that the Newtonian constraint functions as an effective domain-informed regularizer. Although the physical prior is intentionally simplified, the results show that even lightweight inductive biases can enhance robustness and interpretability in financial RL. Future work may incorporate richer financial priors or extend the framework to multi-asset and risk-sensitive settings. Overall, PIKAN provides a practical and principled step toward more stable and economically coherent reinforcement-learning–based portfolio strategies.

\bibliography{DRL_finance}

\end{document}